\documentclass{emulateapj}
\usepackage{epstopdf}

\newcommand{\msun}{M$_{\sun}$}
\newcommand{\ldl}{$\lambda/{\Delta}{\lambda}$}

\newcommand{\teff}{T$_{\rm eff}$}

\newcommand{\ki}{\ion{K}{1}}

\newcommand{\nai}{\ion{Na}{1}}

\newcommand{\lii}{\ion{Li}{1}}

\newcommand{\meth}{CH$_4$}
\newcommand{\wat}{H$_2$O}

\newcommand{\kms}{km~s$^{-1}$}

\newcommand{\vtan}{$V_{\rm tan}$}

\newcommand{\name}{2MASS~J15500845+1455180}
\newcommand{\namesh}{2MASS~J1550+1455}

\slugcomment{Submitted to AJ on 4 August 2009; accepted for publication 16 September 2009}

\shorttitle{L Dwarf Binary 2MASS~J15500845+1455180}
\shortauthors{Burgasser, Dhital \& West}

\begin{document}

\title{Resolved Spectroscopy of M Dwarf/L Dwarf Binaries. III. The ``Wide'' L3.5/L4 Dwarf Binary 2MASS~J15500845+1455180AB}

\author{Adam J.\ Burgasser\altaffilmark{1,2}}
\affil{Center of Astrophysics and Space Sciences, Department of Physics, University of California, San Diego, CA 92093, USA;  aburgasser@ucsd.edu}
\altaffiltext{1}{Also Massachusetts Institute of Technology, Kavli Institute for Astrophysics and Space Research, 77 Massachusetts Avenue, Cambridge, MA 02139, USA}
\altaffiltext{2}{Visiting Astronomer at the Infrared Telescope Facility, which is operated by the University of Hawaii under Cooperative Agreement no. NCC 5-538 with the National Aeronautics and Space Administration, Science Mission Directorate, Planetary Astronomy Program.
}

\author{Saurav Dhital}
\affil{Department of Physics \& Astronomy, Vanderbilt University, Nashville, TN 37235, USA}  

\and

\author{Andrew A.\ West\altaffilmark{1}}
\affil{Institute for Astrophysical Research, Boston University, Boston, MA 02215, USA}

\begin{abstract}
We report the identification of 2MASS~J15500845+1455180 as a 0$\farcs9$ L dwarf
visual binary.  This source is resolved in Sloan Digital Sky Survey (SDSS) images and in near-infrared imaging with the IRTF SpeX imager/spectrometer.  The two components, oriented along a north-south axis, have similar brightnesses in the near-infrared ($\Delta{K}$ $\approx$ 0.2~mag), although 
the fainter northern component is redder in $J-K$ color.  Resolved near-infrared spectroscopy indicates spectral types of 
L3.5 and L4, consistent with its L3 combined-light optical classification based on SDSS data.  Physical association is confirmed through common proper motion, common spectrophotometric distances and low probability of chance alignment.  The projected physical separation of 2MASS~J1550+1455AB, 30$\pm$3~AU at an estimated distance of 33$\pm$3~pc, makes it the widest L dwarf-L dwarf pair identified to date, although such a separation is not unusual among very low-mass field binaries.  The angular separation and spectral composition
of this system makes it an excellent 
target for obtaining a precise lithium depletion age, and a 
potential age standard for low-temperature atmosphere studies.
\end{abstract}


\section{Introduction}

Very low-mass (VLM) stars and brown dwarfs---with masses less than 0.1~{\msun}---are among the most
populous but least understood stellar populations in the Galaxy.  Encompassing the low-mass tail of the stellar/substellar mass function, these faint and low effective temperature ({\teff}) sources are incompletely sampled even in the 
immediate vicinity of the Sun (e.g., \citealt{1995AJ....110.1838R, 2006AJ....132.2360H}).  Over the past decade, wide-field red optical and near-infrared
sky surveys have facilitated the discovery of thousands of new nearby VLM sources, including members
of the recently defined L and T spectral classes \citep{2005ARA&A..43..195K}.  These include roughly 100 VLM multiple systems,\footnote{A current list is maintained at the VLM Binaries Archive, \url{http://www.vlmbinaries.org}.} estimated to comprise 20--25\% of the population
\citep{2006AJ....132..663B, 2007ApJ...668..492A}.  

Studies of VLM multiples have focused in large part on the very tightest
(angular separations $\rho$ $\lesssim$ 0$\farcs$1, projected physical separations $a$ $\lesssim$ 0.1--1~AU)
and the very widest pairs
($\rho$  $\gtrsim$ 1$\arcsec$, $a$ $\gtrsim$ 10--100~AU).
The short periods of the former are useful for Keplerian mass measurements and radius measurements
(e.g., \citealt{2006Natur.440..311S}), both required to test low-mass stellar structure models (e.g., \citealt{2009AIPC.1094..102C}).   For those systems
with independent age determinations (cluster members, companions to age-dated stars),
these measurements also serve to test evolutionary theory \citep{2004ApJ...615..958Z, 2009ApJ...692..729D}.  
The identification
and study of tight VLM multiples is challenged, however, by the need for adaptive optics-enhanced and/or space-based
high-resolution imaging, and in a few cases large-aperture high resolution spectroscopic monitoring
\citep{2002AJ....124..519R, 2006AJ....132..663B, 2007ApJ...666L.113J}.

In contrast, wide VLM multiples have prohibitively long orbits for mass measurements, but detailed investigations of their coeval components can be made.  These systems facilitate comparative analyses of low-temperature atmospheres and magnetic emission in the absence of age and composition dependencies (e.g., \citealt{2006AJ....131.1007B, 2006AJ....132.2074M, 2006A&A...456..253M, 2009ApJ...695..788K}).  Wide VLM pairs are also relatively rare, with systems wider than $a$ $\gtrsim$ 20~AU comprising $\lesssim$2\% of all VLM multiples in the field (\citealt{2007ApJ...668..492A,2007ApJ...667..520C}), although 
a few systems exceeding 1000~AU separation have
been identified (e.g., \citealt{2007A&A...462L..61C, 2007ApJ...659L..49A, 2009ApJ...698..405R}).  
The rarity of wide VLM binaries in the field and in some young clusters (e.g., \citealt{2009arXiv0908.1385K}) may be a consequence of formation mechanisms unique to VLM stars and brown dwarfs (e.g., embryonic dynamical ejection: \citealt{2001AJ....122..432R, 2002MNRAS.332L..65B}; fragmentation of circumstellar disks: \citealt{2001ApJ...551L.167B, 2004AJ....127..455J}) or dynamical evolution within and/or outside the natal cloud
(e.g., \citealt{2007ApJ...660.1492C}).  In addition, wide VLM pairs are often found to be triples or quadruples (e.g,. \citealt{2006ApJ...645L.153P}; S.\ Dhital et al.\ 2009, in preparation), enabling studies of higher order multiplicity associated with star and brown dwarf formation.

In this article, we report the discovery of a new wide L dwarf pair, {\name} (hereafter {\namesh}; \citealt{2007AJ....133..439C}) identified by direct imaging and resolved spectroscopy to be a 0$\farcs$9, near-equal brightness binary.  
In Section~2 we describe imaging and spectroscopic observations of the source, and ascertain the separation, relative near-infrared magnitudes and individual spectral classifications of its components.  In Section~3 we review evidence that this source is a physical binary based on constraints in common proper motion, common spectrophotometric distances and low probability of chance alignment.  The properties of {\namesh} are discussed in the context
of other VLM binaries in Section~4, along with motivation to measure its age using the binary brown dwarf {\lii} test.
Results are summarized in Section~5.

\section{Observations}

\subsection{Target Properties and SDSS Measurements}

{\namesh} was originally identified by \citet{2007AJ....133..439C} in a color- and magnitude-selected search of the Two Micron All Sky Survey (2MASS; \citealt{2006AJ....131.1163S}) for nearby late-type M and L dwarfs.  It was selected as a near-infrared bright ($K_s$ = 13.26$\pm$0.04~mag) but optically faint source with red near-infrared colors ($J-K_s$ = 1.52$\pm$0.05~mag).  The epoch 2009 February 6 (UT) 2MASS images show an unresolved point source.
\citet{2007AJ....133..439C} reported an optical spectral classification of L2: on the \citet{1999ApJ...519..802K} scale, the uncertainty arising from low signal-to-noise data (I.~N.\ Reid \& K.\ L.\ Cruz, 2009, private communication).  {\namesh} has a relatively small proper motion, $\mu$ = 0.165$\pm$0.015~$\arcsec$~yr$^{-1}$ \citep{2008MNRAS.384.1399J}, which combined with the 31~pc distance estimate of \citet{2007AJ....133..439C} yields an estimated tangential velocity {\vtan} = 24~{\kms}, typical for a ``normal'' thin disk L dwarf (e.g., \citealt{2009AJ....137....1F}).

\begin{figure}
\centering
\epsscale{1.0}
\plotone{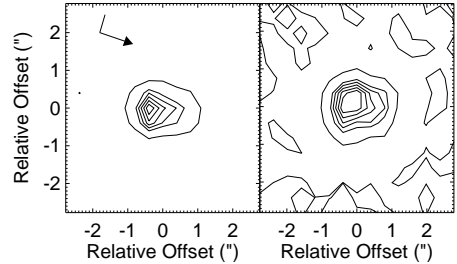}
\caption{Contour plots of flux in epoch 2005 March 10 (UT) SDSS $i$- (left) and $z$-band images (right) at the position of {\namesh}.  Contours of  0.98, 0.96, 0.94, 0.92, 0.9, 0.85 and 0.8 times the peak flux of the central source are shown. The axis scale is arcseconds on the sky relative to the SDSS coordinates.  The orientation of the images is indicated by the arrow (head to north, foot to east).
\label{fig_sdssimage}}
\end{figure}

\begin{figure}
\centering
\epsscale{1.0}
\plotone{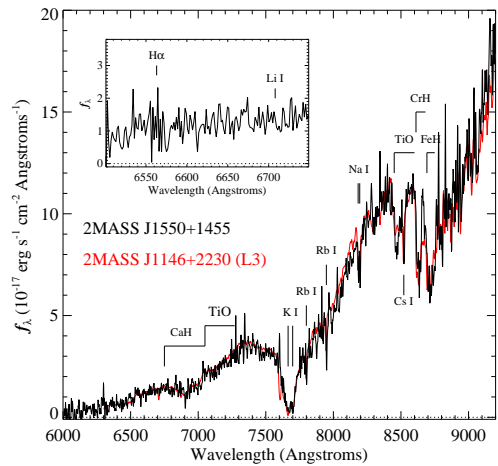}
\caption{SDSS spectrum of {\namesh} (thin black line) compared to the 
L3 spectral standard 2MASS~J1146345+223053 (thick red line; \citealt{1999ApJ...519..802K}).  Both spectra are smoothed to a common resolution of {\ldl} = 1500, and the standard spectrum is scaled to match the SDSS spectrum at 8400~{\AA}.   Key spectral features are labeled.  Inset box shows the region around H$\alpha$ (6563~{\AA}) and {\lii} lines (6708~{\AA}) in the unsmoothed SDSS spectrum of {\namesh}. \label{fig_optspec}}
\end{figure}

{\namesh} was also imaged and targeted for spectroscopy by the Sloan Digital Sky Survey (SDSS; \citealt{2000AJ....120.1579Y}).  This source is morphologically classified in SDSS as a galaxy due to its extended point spread function (PSF), as shown in Figure~\ref{fig_sdssimage}.  Both $i$- and $z$-band images show a faint extension toward the north, which as discussed below is consistent with a faint, marginally resolved companion.
Spectroscopic data from SDSS Data Release 7 (DR7; \citealt{2009ApJS..182..543A}) are shown in Figure~\ref{fig_optspec}.  These data are superior to those in the \citet{2007AJ....133..439C} study, and comparison to the \citet{1999ApJ...519..802K} spectral standards indicate an optical classification of L3 (an uncertainty of 0.5 subtypes is assumed).
There is no indication of either H$\alpha$ emission (6563~{\AA}) or {\lii} absorption (6708~{\AA})
to limiting pseudo-equivalent widths\footnote{Equivalent widths relative to the local continuum.} of --1.8~{\AA} and 1.2~{\AA}, respectively.
We infer a radial velocity for {\namesh} of $-$7$\pm$12~{\kms} based on cross-correlation with similarly-typed 
L dwarf velocity templates also observed by SDSS (S.\ Schmidt et al., in preparation).

\subsection{Imaging Observations}

{\namesh} was targeted with the 3~m NASA Infrared Telescope Facility (IRTF)
SpeX spectrograph \citep{2003PASP..115..362R} on 2009 June 29 (UT), as part of a program
to identify unresolved L dwarf/T dwarf spectral binaries (e.g., \citealt{2008ApJ...681..579B}).
Conditions were clear with excellent seeing of 0$\farcs$4--0$\farcs$5 at $J$- and $K$-bands.
Acquisition images obtained with SpeX's imaging channel revealed two point sources
at the position of {\namesh} aligned along a north-south axis, in the same orientation as the SDSS extended PSF.  Four 20~s dithered exposures
were obtained at an airmass of 1.31 
in each of the MKO\footnote{Mauna Kea Observatory filter system; see \citet{2002PASP..114..180T} and \citet{2002PASP..114..169S}.} $J$, $H$, and $K$ filters.  
We also obtained two 20~s dithered images of the nearby ($\rho \sim$ 9.4$\arcmin$) point source 2MASS~J15504059+1500305
($K_s$ = 13.11$\pm$0.03~mag, $J-K_s$ = 0.73$\pm$0.04~mag) in each of the $JHK$ filters immediately after
the {\namesh} observation 
to serve as a PSF calibrator.

\begin{figure}
\centering
\epsscale{1.0}
\plotone{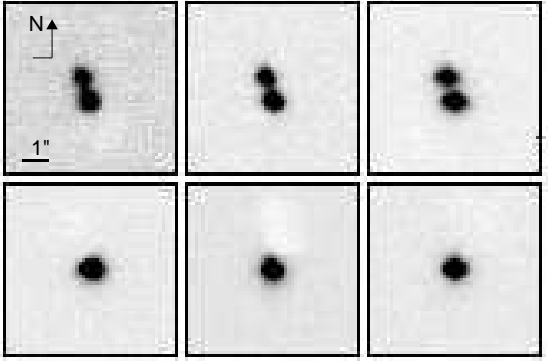}
\caption{SpeX images of {\namesh}AB (top) and the nearby point source 2MASS~J15504059+1500305 (bottom)
in the $J$-, $H$- and $K$-bands (left to right).  The images show 
6$\farcs$12$\times$6$\farcs$12 (51$\times$51 pixels) regions around both sources, 
aligned with north up and east to the left.
\label{fig_image}}
\end{figure}

Imaging data were reduced in a standard manner using custom Interactive Data Language (IDL) routines.
After mirror-flipping the raw imaging data along the y-axis to reproduce the sky orientation
(J.~T.\ Rayner, 2009, private communication),
sky images for each filter were produced by median-combining all of the {\namesh} and PSF
calibrator images.  These were subtracted from the raw imaging data, and each sky
image was also normalized and used as a flat-field frame to correct for pixel-to-pixel 
response variations.  Subsections of each image, 51 pixels (6$\farcs$12) on a side and centered on the target source, were extracted from these calibrated frames.  A final image for each filter/target pair 
was produced by averaging the registered subframes together, rejecting 5$\sigma$ pixel outliers for
the {\namesh} observations.   These combined images are shown in Figure~\ref{fig_image}.  
The two components of {\namesh} are clearly 
resolved, separated by roughly 1$\arcsec$ along a nearly north-south axis, the southern
component appearing to be slightly brighter in all three filter bands.  Hereafter, we refer to this
component as {\namesh}A and the northern component as {\namesh}B.

\subsection{PSF Fitting}

Component magnitudes and the angular separation of the {\namesh} pair were determined by PSF fits to the reduced imaging data, following the prescription described in \citet{2006AJ....132.2074M}.
Fits were made to each individual image frame using both PSF calibrator images in a given filter band, for a total of eight independent measures of the relative $JHK$ component magnitudes
and 24 independent measures of the separation and orientation of the pair.    
Measurements of the latter
were converted from pixels to arcseconds assuming a 
plate scale of 0$\farcs$120$\pm$0$\farcs$002 pixel$^{-1}$ (J.\ T.\ Rayner, 2005, private communication) and no distortion.
The position angle was assumed to be accurate to within 0$\fdg$25 (ibid.).

To test for systematic effects in the derived relative magnitudes, we performed the same fits on synthetic binary images constructed from the PSF
frames.  For each filter, synthetic images were made by combining two randomly selected calibrator images, one shifted according to the measured separations (with an additional random shift based on the separation uncertainties) and scaled by flux ratios spanning 0.2 to 1.0 (magnitude differences of 1.75 to 0~mag).
A linear fit between input and output flux ratios for 200 trials indicates systematic shifts of 3-7\% depending on the filter band (largest at $K$), which were incorporated into the reported photometry and uncertainties.

\begin{deluxetable}{lc}
\tabletypesize{\footnotesize}
\tablecaption{Results of PSF Fitting\label{tab_psf}}
\tablewidth{0pt}
\tablehead{
\colhead{Parameter} &
\colhead{Value} \\
}
\startdata
\cline{1-2}
\multicolumn{2}{c}{SDSS Epoch 2005 March 10 (UT)} \\
\cline{1-2}
$\Delta{\alpha}\cos{\delta}$ ($\arcsec$)\tablenotemark{a} & 0.24$\pm$0.04   \\
$\Delta{\delta}$ ($\arcsec$)\tablenotemark{a} & 0.92$\pm$0.06  \\
$\rho$ ($\arcsec$) & 0.95$\pm$0.06   \\
$\theta$ ($\degr$)\tablenotemark{a} & 14.5$\pm$1.3   \\
$\Delta{i}$ (mag) & 0.8$\pm$0.3 \\
$\Delta{z}$ (mag) & 0.7$\pm$0.3 \\
\cline{1-2}
\multicolumn{2}{c}{SpeX Epoch 2009 June 29 (UT)} \\
\cline{1-2}
$\Delta{\alpha}\cos{\delta}$ ($\arcsec$)\tablenotemark{a} & 0.26$\pm$0.02   \\
$\Delta{\delta}$ ($\arcsec$)\tablenotemark{a} & 0.87$\pm$0.03   \\
$\rho$ ($\arcsec$) & 0.91$\pm$0.03   \\
$\theta$ ($\degr$)\tablenotemark{a} & 16.6$\pm$1.3   \\
$\Delta{J}$ (mag) & 0.41$\pm$0.03 \\
$\Delta{H}$ (mag) & 0.29$\pm$0.05 \\
$\Delta{K}$ (mag) & 0.23$\pm$0.03 \\
\enddata
\tablenotetext{a}{Angular separation ($\rho$) and position angle ($\theta$) measured from the brighter primary to the fainter secondary.}
\end{deluxetable}

Results are listed in Table~\ref{tab_psf}.  The angular separation of the pair is inferred to be 
0$\farcs$91$\pm$0$\farcs$03 at a position angle of 16$\fdg$6$\pm$1$\fdg$1 (east of north, vector pointing from primary to secondary).
The magnitude differences of the two components
decrease from $J$ to $K$, indicating a secondary that is significantly
redder than the primary, $J-K_s$ = 1.63$\pm$0.06~mag versus 1.43$\pm$0.06~mag.\footnote{These colors are on the 2MASS photometric system, based on the MKO to 2MASS photometric conversions described in Section~3.} 
As described below, this is consistent with the slightly later
spectral classification inferred for this component.

We performed the same analysis on the epoch 2005 March 10 (UT) SDSS $i$ and $z$ images, using three nearby point sources as PSF calibrators.  The coarser plate scale of the SDSS images (0$\farcs$396 pixel$^{-1}$; \citealt{2003AJ....125.1559P}) and larger PSFs result in both larger astrometric uncertainties 
and larger systematic offsets in the relative photometry (up to 30\%, based on simulations equivalent to those described above).  Nevertheless, the inferred separation and orientation of the components formally agree with the SpeX results.  There is also an indication that the northern component is considerably fainter in the SDSS bands, $\Delta{i}$ = 0.8$\pm$0.3~mag and $\Delta{z}$ = 0.7$\pm$0.3~mag.

\subsection{Spectroscopic Observations}

Both components of {\namesh} were also observed on 2009 June 29 (UT) with the  
prism-dispersed mode of SpeX,
which provides 0.75--2.5~$\micron$ continuous
spectroscopy with resolution {\ldl} $\approx 120$ for the 0$\farcs$5
slit employed (dispersion across the chip is 20--30~{\AA}~pixel$^{-1}$).
The components were observed simultaneously by aligning the slit to 
a position angle of 15$\degr$; note that this is 62$\degr$ off from the parallactic angle, so 
some wavelength-dependent light loss is expected (although it was somewhat
mitigated by the excellent seeing; see \citealt{2007ApJ...658..557B}).
Six exposures of 120~s each were obtained over an airmass ranging from 1.25 to 1.29,
dithering in an ABBA pattern along the slit. 
Guiding was performed on spillover light from the
brighter component, imaged through the H+K notch filter on the SpeX imaging channel.
For telluric absorption correction and flux calibration, we
observed the nearby ($\rho$ = 7$\fdg$1) A0 V star HD 136831 within an hour of the {\namesh} observation,
at a similar airmass (1.20) and using the same instrument configuration, 
but with the slit aligned along the parallactic angle.
Data were reduced using the SpeXtool package, version 3.3
\citep{2003PASP..115..389V, 2004PASP..116..362C} with standard settings;
see \citet{2007ApJ...658..557B} for details.  We took care in the extraction step
to select only the opposite wings of the spatial PSF profiles of each
component in order to minimize flux contamination.  Based on the width of these
profiles (full width at half maximum of $\sim$0$\farcs$5), the separation of the components (0$\farcs$9; see below)
and their similar brightnesses, we estimate no more than 5--10\% contamination in either
spectrum over the 1.0--2.4~$\micron$ range.

\begin{figure}
\centering
\epsscale{1.0}
\plotone{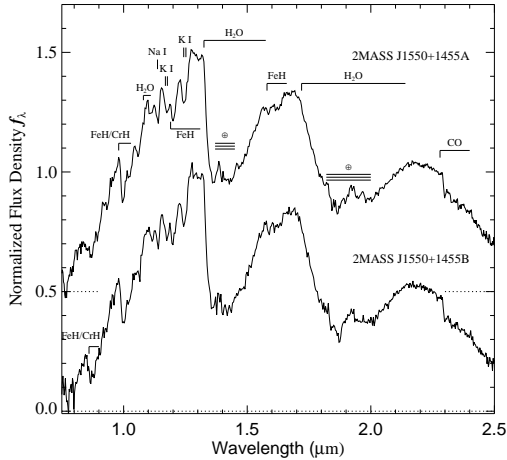}
\caption{Near-infrared spectra of {\namesh}A (top) and {\namesh}B (bottom) obtained with IRTF/SpeX.
Data are normalized at the 1.3~$\micron$ spectral peak, with the spectrum of  {\namesh}A
offset by a constant for clarity (dotted lines).  Near-infrared spectral features characteristic of L dwarfs are labeled.
\label{fig_nirspec}}
\end{figure}

The reduced spectra for both components are shown in Figure~\ref{fig_nirspec}.
Both exhibit classic signatures of L dwarf near-infrared spectra: 
deep {\wat} absorption bands centered at 1.4 and 1.9~$\micron$;
strong CO absorption at 2.3~$\micron$;
strong FeH absorption at 0.86, 0.99 and 1.6~$\micron$;
unresolved {\ki} and {\nai} doublets at 1.14, 1.17 and 1.25~$\micron$;
no detectable {\nai} absorption at 2.2~$\micron$ (absent in L2 and later dwarfs; \citealt{2003ApJ...596..561M});
and a steep 0.8--1.2~$\micron$ spectral slope.
Both spectra appear to be representative of normal L dwarfs, lacking the extremely red or extremely
blue spectral energy distributions found in young/dusty and old/metal-poor L dwarfs,
respectively (e.g., \citealt{2003ApJ...592.1186B, 2007ApJ...657..511A, 2008ApJ...686..528L}).
{\wat} and FeH absorption is somewhat deeper in the fainter component; note in particular
the depths of the 1.3~$\micron$ and 0.86~$\micron$ bands, respectively.

\begin{deluxetable}{lccccl}
\tabletypesize{\footnotesize}
\tablecaption{Near Infrared Spectral Indices\label{tab_indices}}
\tablewidth{0pt}
\tablehead{
&
\multicolumn{2}{c}{{\namesh}A} & 
\multicolumn{2}{c}{{\namesh}B} &
 \\ 
\cline{2-3} \cline{4-5}
\colhead{Index}  & 
\colhead{Value} &
\colhead{SpT} &
\colhead{Value} &
\colhead{SpT} & 
\colhead{Reference} \\
}
\startdata
{\wat}-J & 0.827 & L3.5 & 0.786 & L4.5 &  1,2 \\
{\wat}-H & 0.776 & L3.5 & 0.739 & L5 &  1,2 \\
{\meth}-K & 1.064 & L2\tablenotemark{a} & 1.040 & L4 &  1,2 \\
{\wat}-A & 0.634 & L3 & 0.607 & L4 & 3 \\
{\wat}-B & 0.691 & L3.5 & 0.663 & L4 & 3 \\ 
K1 & 0.268 & L3 & 0.297 & L3.5 &  3,4 \\
{\wat}-1.5~$\micron$ & 1.437 & L3  &  1.526 & L4 & 5 \\ 
{\meth}-2.2~$\micron$ & 0.940 & L3/L4  & 0.962 & L4 &  5 \\
\enddata
\tablenotetext{a}{Not used to compute final (mean) classification of this source.}
\tablerefs{(1) \citet{2006ApJ...637.1067B}; (2) \citet{2007ApJ...659..655B}; (3) \citet{2001AJ....121.1710R}; (4) \citet{1999AJ....117.1010T}; (5) \citet{2002ApJ...564..466G}.}
\end{deluxetable}

Near-infrared classifications for each spectrum were derived using a suite of spectral indices
and spectral type/index relations defined by \citet{1999AJ....117.1010T, 2001AJ....121.1710R, 2002ApJ...564..466G, 2006ApJ...637.1067B}; and \citet{2007ApJ...659..655B}.
Table~\ref{tab_indices} lists the values of eight indices sampling the prominent {\wat} bands
and shape of the $K$-band peak.  The subtypes inferred from these indices are nearly all in 
agreement, with the exception of the {\meth}-K index for {\namesh}A which is not particularly sensitive to early-L subtype.  Ignoring this index,
we infer average types of L3.5 and L4 for {\namesh}A and {\namesh}B, respectively, 
with scatter of 0.3--0.4~subtypes.  These classifications are formally consistent with the
L3 combined-light optical classification of the system, particularly as the 
secondary is considerably fainter than the primary at these wavelengths.
Note that each index consistently classifies the B component 0.5-1.5 subtypes later than the A component.

The near-infrared types of {\namesh}A and {\namesh}B correspond to effective temperatures {\teff} $\approx$ 1910$\pm$110~K and 1840$\pm$110~K, 
respectively, according to the {\teff}/spectral type relation of \citet[uncertainties include classification error and scatter in relation]{2008ApJ...685.1183L}.
Estimates of the component masses for ages of 0.5, 1 and 5~Gyr based on these temperatures and the evolutionary models of \citet{1997ApJ...491..856B} are listed in Table~\ref{tab_properties}; estimate are roughly 10\% smaller for the DUSTY and COND evolutionary models of \citet{2000ARA&A..38..337C} and \citet{2003A&A...402..701B}.
Note that the near-unity mass ratio of this system, $q \equiv$ M$_2$/M$_1$ $>$ 0.95 for ages of 0.5--5~Gyr, is typical of the majority of VLM binaries \citep{2007ApJ...668..492A}.

\section{Is {\namesh} A Physical Binary?}

The presence of two similarly-typed, near equal-brightness
L dwarfs at the position of {\namesh} strongly suggests a physically bound pair, and we examined three lines of evidence for association: common proper motion, common
distance, and low probability of chance alignment or random pairing.    For the first criterion, the detection of the secondary in both SDSS and SpeX images provides a stringent constraint on the relative motion of the two sources over the 
intervening 4.3~yr. Assuming that the component alignment is the same in both epochs (i.e., that the northern component is the same star in both images), we infer relative motions of 4$\pm$10~mas~yr$^{-1}$ in Right Ascension and --12$\pm$15~mas~yr$^{-1}$ in declination; i.e., consistent with no relative motion.  These limits correspond to a relative tangential velocity
of {\vtan} $\leq$ 3~{\kms} at the adopted distance of {\namesh} (see below), of order their
estimated orbital velocities ($\sim$1~{\kms}; see below).
We therefore rule out significant differential motion between the components of this system.

For the second criterion, distances were computed from the individual 
component near-infrared magnitudes, using the combined light 2MASS photometry and
relative photometry from our SpeX measurements; see Table~\ref{tab_properties}.\footnote{We included small corrections to the relative magnitudes in converting from MKO to 2MASS
photometric systems: 0.009, -0.006 and -0.003~mag in the $J$-, $H$- and $K$/$K_s$-bands
respectively, calculated directly from the spectral data.  See also \citet{2004PASP..116....9S}.} These magnitudes, combined with the $M_J$/spectral type relation of \citet{2003AJ....126.2421C}, indicate
spectrophotometric distance estimates of 31$\pm$3~pc for the primary
and 34$\pm$3~pc for the secondary, where the uncertainties include both photometric
uncertainties and scatter in the Cruz et al.\ relation.  These distances agree to within the uncertainties and 
are consistent with an average distance of 
33$\pm$3~pc.  

Finally, we calculated the probability of
chance alignment based on the three dimensional position of {\namesh} in
the Galaxy. S.\ Dhital et al.\ (2009, in preparation) have constructed a Galactic model, constrained with empirical
stellar density profiles \citep{2008ApJ...673..864J, Bochanski2009} and disk velocity
distributions \citep{2007AJ....134.2418B}, to assess the physical association of wide binary
systems. The Monte Carlo model recreates a random distribution of stars in an
1800$\arcsec \times$ 1800$\arcsec$
region of sky, centered on the coordinates of a given binary system, and out to a
heliocentric distance of 2500~pc. As all the simulated stars are non-associated, the number of stars
found within the separation of the binary is a measure of the number of chance
alignments expected at that position in the Galaxy. In $10^7$
realizations, we found the probability of chance alignment on the sky
to be 0.27\%.  Importantly, none of the projected matches were within the range of spectrophotometric
distances estimated for the {\namesh} components, corresponding to a $\lesssim$ 10$^{-7}$ probability of positional coincidence. 

Based on these various lines of evidence, we conclude that {\namesh}
is a physically bound L dwarf binary.

\section{Discussion}

For an estimated distance of 33$\pm$3~pc, we infer a projected separation
of 30$\pm$3~AU for the components of {\namesh}, making it the widest L dwarf-L dwarf pair identified to date.  It is $\sim$20\% wider in physical separation than 
the L1.5+L4.5 pair 2MASS~J1520022-442242AB ($a$ = 22$\pm$2~AU; \citealt{2007ApJ...658..557B, 2007MNRAS.374..445K}), although that system has a wider angular separation (1$\farcs$174$\pm$0$\farcs$016).  Based on its separation and estimated component masses (Table~\ref{tab_properties}), we estimate a minimum orbital period $P$ $\gtrsim$ 400--500~yr.

The extremity of its separation does not necessarily make {\namesh} an unusually wide VLM pair.
\citet{2003ApJ...586..512B} quantified an empirical relation between the maximum separation limit for VLM pairs in the field and their total mass: $a_{max}$ (AU) = 1400 (M$_{tot}$/M$_{\sun}$)$^2$, applicable for systems with M$_{tot}$ $\lesssim$ 0.2~{\msun}.  The projected separation of {\namesh} is within a factor of two of this limit, $a_{max}$ = 16--35~AU for ages of 0.5--5~Gyr based on the estimated masses listed in Table~\ref{tab_properties}.  The projected binding energy of this system, $-U_g \equiv$ GM$_1$M$_2$/$a \approx$ (2--4)$\times$10$^{42}$~erg, is also near empirical limits encompassing the majority of VLM field systems \citep{2003ApJ...587..407C, 2007ApJ...660.1492C}.   
In contrast, a handful of slightly more massive late-type M dwarf/L dwarf VLM pairs have been identified
in the field 
that break these empirical limits, with separations exceeding 1000~AU \citep{2007A&A...462L..61C, 2007ApJ...659L..49A, 2009ApJ...698..405R}; there have also been found comparably wide, weakly bound low-mass brown dwarf pairs in nearby young star forming regions and associations (e.g., \citealt{2004ApJ...614..398L, 2006A&A...460..635C, 2006Sci...313.1279J, 2007ApJ...660.1492C, 2008ApJ...673L.185B}; however, see \citealt{2009arXiv0908.1385K}).
In the context of these systems, it is clear that {\namesh} is in no sense an unusually wide system.

What makes {\namesh} potentially interesting is the fact
that both components have estimated masses near the lithium burning minimum mass limit, 
$\sim$0.06~{\msun} for solar metallicity \citep{1997A&A...327.1039C, 2001RvMP...73..719B}.  This makes it a potential target for applying the binary lithium age-dating technique first proposed
by \citet{2005ApJ...634..616L} for VLM pairs.  Specifically, if the 6708~{\AA} {\lii} absorption line is present in the spectrum of the secondary of a coeval VLM pair (indicating a mass $\lesssim$ 0.06~{\msun} and a lower age bound) but not the primary (indicating a mass $\gtrsim$ 0.06~{\msun} and an upper age bound), one obtains a finite constraint on the system age.\footnote{This age-dating technique parallels the lithium depletion boundary technique used to date 10--200~Myr-old clusters (e.g., \citealt{1993ApJ...404L..17M, 1997ApJ...482..442B, 1998ApJ...499L.199S}), but is applicable for older field systems stradding the lithium burning minimum mass.}
The absence of the {\lii} line in the combined-light optical spectrum (Figure~\ref{fig_optspec}) is in fact promising, as it indicates that lithium is depleted in the atmosphere of the primary.  This places an age constraint of $\tau$ $\gtrsim$ 0.55~Gyr based on the estimated {\teff} of this component (including uncertainty) and the evolutionary models of \citet{1997ApJ...491..856B}.  The lack of a clear {\lii} line may also indicate that lithium is depleted in the secondary as well; alternately, the line may be obscured by the relatively bright continuum from the primary.  Using as a template spectral data for the L3 spectral standard 2MASS~J1146345+223053, a source which exhibits {\lii} absorption (\citealt{1999ApJ...519..802K}, Figure~\ref{fig_optspec}), and assuming a $\sim$1~mag brightness difference in the 6708~{\AA} region, we determined that a signal-to-noise ratio of 20 or better is required to detect {\lii} absorption from the secondary alone.  In contrast, SDSS data for {\namesh} have a signal-to-noise of roughly 5 in this region.  Hence, we cannot rule out the  presence of the {\lii} line in the secondary.  If this line is present, it would indicate a bounded age constraint of 0.55 $\lesssim$ $\tau$ $\lesssim$ 0.8~Gyr for the system (the corresponding constraint based on the \citealt{2003A&A...402..701B} evolutionary models are 0.6 $\lesssim$ $\tau$ $\lesssim$ 1.1~Gyr).

Such a young age is not inconsistent with
the kinematics of {\namesh}.  The inferred space velocities,
[$U, V, W$] = [26$\pm$6, 0$\pm$4, --14$\pm$7]~{\kms} evaluated in the local standard of rest \citep{1998MNRAS.298..387D}, are nominally consistent with that of a young, thin disk system.  
The $J-K_s$ colors of the primary is slightly blue for its inferred type, since
$\langle{J-K_s}\rangle$ $\approx$ 1.6--1.7~mag for L3 dwarfs \citep{2000AJ....120..447K, 2009AJ....137....1F}.
This could be an indication of an older age, since unusually blue near-infrared colors for VLM dwarfs have been associated with both older kinematics and higher surface gravities (e.g., \citealt{2004AJ....127.3553K, 2009AJ....137....1F}).
However, variations in cloud properties can also produce variations in near-infrared color, an effect most pronounced in mid-type L  dwarfs \citep{2006AJ....131.2722C, 2008ApJ...674..451B}.
Resolved optical spectroscopy would settle this issue, and could identify {\namesh} as
an important age standard for VLM studies.

\section{Summary}

We have resolved the L dwarf {\namesh} into a 0$\farcs$9 visual binary with L3.5 and L4 components.   Evidence based on the common proper motion, common spectrophotometric distance and low probability of random alignment confirms this
system as a physically bound pair.  While {\namesh} has the widest projected physical separation of any L dwarf-L dwarf binary identified to date, 30$\pm$3~AU at an estimated distance of 33$\pm$3~pc, it is not unusual for VLM multiples in the field.  The effective temperatures of its components and lack of {\lii} absorption in its combined light spectrum identify {\namesh} as a good candidate for determining a precise systemic age through the binary lithium age-dating technique
of \citet{2005ApJ...634..616L}, necessitating resolved optical spectroscopy that is eminently feasible for this well-separated pair.

\acknowledgments

The authors would like to thank telescope operator Paul Sears
and instrument specialists John Rayner
for their assistance during the IRTF observations,
I.\ N.\ Reid and K.\ L.\ Cruz for their helpful input in the preparation of the manuscript,
and J.\ A.\ Caballero for his thorough and expert review.
This publication makes use of data 
from the Two Micron All Sky Survey, which is a
joint project of the University of Massachusetts and the Infrared
Processing and Analysis Center, and funded by the National
Aeronautics and Space Administration and the National Science Foundation.
2MASS data were obtained from the NASA/IPAC Infrared
Science Archive, which is operated by the Jet Propulsion
Laboratory, California Institute of Technology, under contract
with the National Aeronautics and Space Administration.
Funding for the SDSS and SDSS-II has been provided by the Alfred P. Sloan Foundation, the Participating Institutions, the National Science Foundation, the U.S. Department of Energy, the National Aeronautics and Space Administration, the Japanese Monbukagakusho, the Max Planck Society, and the Higher Education Funding Council for England. 
The SDSS is managed by the Astrophysical Research Consortium for the participating institutions 
(\url{http://www.sdss.org}).
This research has made use of the SIMBAD database,
operated at CDS, Strasbourg, France; 
the Very-Low-Mass Binaries Archive housed at 
\url{http://www.vlmbinaries.org} and maintained by Nick Siegler, Chris Gelino, and Adam Burgasser;
and the M, L, and T dwarf compendium housed at DwarfArchives.org and 
maintained by Chris Gelino, Davy Kirkpatrick, and Adam Burgasser. 
The authors wish to recognize and acknowledge the 
very significant cultural role and reverence that 
the summit of Mauna Kea has always had within the 
indigenous Hawaiian community.  We are most fortunate 
to have the opportunity to conduct observations from this mountain.

Facilities: \facility{IRTF (SpeX)}


\begin{deluxetable}{lcc}
\tabletypesize{\footnotesize}
\tablecaption{Properties of {\name}AB \label{tab_properties}}
\tablewidth{0pt}
\tablehead{
\colhead{Parameter} &
\colhead{{\namesh}A} &
\colhead{{\namesh}B} \\
}
\startdata
Near Infrared Spectral Type & L3.5 & L4 \\
$i$ (mag) & 19.75$\pm$0.09 & 20.56$\pm$0.19 \\ 
$z$ (mag) & 17.95$\pm$0.09 & 18.68$\pm$0.17 \\ 
$J$ (mag)  & 15.34$\pm$0.04 & 15.76$\pm$0.05 \\ 
$H$ (mag)  & 14.42$\pm$0.05 & 14.70$\pm$0.05 \\ 
$K_s$ (mag) & 13.91$\pm$0.04 & 14.13$\pm$0.04 \\ 
$i-z$ (mag) & 1.80$\pm$0.13 & 1.88$\pm$0.25 \\
$J-K_s$ (mag) & 1.43$\pm$0.06 & 1.63$\pm$0.06 \\
{\teff}\tablenotemark{a} (K) & 1910$\pm$110 & 1840$\pm$110 \\
Mass at 0.5~Gyr\tablenotemark{b} ({\msun}) & 0.056 & 0.054 \\
{Mass at}~1~Gyr\tablenotemark{b} ({\msun}) & 0.070 & 0.067 \\
{Mass at}~5~Gyr\tablenotemark{b} ({\msun}) & 0.079 & 0.078 \\
Distance\tablenotemark{c} (pc) & 31$\pm$3 & 34$\pm$3 \\ 
\cline{1-3}
 &  \multicolumn{2}{c}{{\namesh}AB} \\
\cline{1-3}
Optical Spectral Type & \multicolumn{2}{c}{L3} \\
Distance (pc) & \multicolumn{2}{c}{33$\pm$3} \\
Projected Separation (AU) & \multicolumn{2}{c}{30$\pm$3} \\
{\vtan} ({\kms}) & \multicolumn{2}{c}{26$\pm$3} \\
$V_{\rm rad}$ ({\kms}) & \multicolumn{2}{c}{-7$\pm$12} \\
$U, V, W$ ({\kms}) & \multicolumn{2}{c}{26$\pm$6, 0$\pm$4, -14$\pm$7} \\
\enddata
\tablenotetext{a}{Estimate based on the {\teff}/spectral type relation of \citet{2008ApJ...685.1183L} and including uncertainties in relation and component classifications ($\pm$0.3--0.4~subtypes).}
\tablenotetext{b}{Estimates based on the evolutionary models of \citet{1997ApJ...491..856B}.}
\tablenotetext{c}{Estimate based on the $M_J$/spectral type relation of \citet{2003AJ....126.2421C}; uncertainties include photometric uncertainties and scatter in the Cruz et al.\ relation.}
\end{deluxetable}

\end{document}